\documentclass[12 pt]{article} 
\usepackage{times}
\usepackage{authblk,float}
\usepackage{graphicx}
\usepackage{bm}
\usepackage[pagebackref=false,colorlinks,linkcolor=blue,citecolor=blue]{hyperref}
\usepackage[margin=1.1 in]{geometry}
\usepackage{booktabs} 
\usepackage{array} 
\usepackage{paralist} 
\usepackage{verbatim} 
\usepackage{subfig} 
\usepackage{fancyhdr} 
\usepackage{sectsty}
\usepackage[nottoc,notlof,notlot]{tocbibind} 
\usepackage[titles,subfigure]{tocloft} 
\usepackage{cite}
\usepackage{amsmath}
\usepackage{amsfonts}
\usepackage{amssymb}
\usepackage[usenames, dvipsnames]{color}
\usepackage{float}
\usepackage{xcolor}
\usepackage{tikz,pgfplots}
\begin{document}

\title{ \textmd{
The  DKP oscillator in spinning  cosmic string background  
}}

\author{\large \textit {Mansoureh Hosseinpour}$^{\ 1}$\footnote{E-mail:hosseinpour.mansoureh@gmail.com(corresponding author)} and~\textit{Hassan Hassanabadi}$^{\ 1}$\footnote{E-mail: h.hasanabadi@shahroodut.ac.ir}
	\\
	
	\small \textit {$^{\ 1}$ Faculty of Physics, Shahrood University of Technology, Shahrood, Iran}\\
	\small \textit{P.O. Box 3619995161-316} }

\date{}

\maketitle

\begin{abstract}
  \textmd{In this article, we investigate the behavior of relativistic spin-zero bosons in the space-time generated by a spinning cosmic string. We obtain the generalized beta-matrices in terms of the flat spacetime ones and rewrite the covariant form of Duffin-Kemmer-Petiau (DKP) equation in spinning cosmic string space time.  We find the solution of DKP oscillator and determine the energy levels. We also discuss the influence of the topology of the cosmic string on the energy levels and the DKP spinors.
}
\end{abstract}

\begin{small}
\textmd{\textbf{Key Words}:  DKP equation, cosmic string, bosons, DKP oscillator , curved space-time.}
\end{small}
\\\\
\begin{small}
\textmd{\textbf{PACS}: 03.65.Pm, 04.62. +v, 04.20.Jb  }

\end{small}

\section{Introduction}
The Duffin-Kemmer-Petiau (DKP) equation has been used to describe relativistic spin-0 and spin-1 bosons \cite{1,2,3,4}. The DKP equation has five and ten dimensional representation respectively for spin-0 and spin-1 bosons\cite{5}. This equation is compared to the Dirac equation for fermions \cite{6}. The DKP equation has been widely investigated in many areas of physics. 
The DKP equation has been investigated in the momentum space with the presence of minimal length \cite{7,8} and for spins 0 and 1 in a noncommutative space \cite{9,10,11,12}.  Also, the DKP oscillator has been studied in the presence of topological defects \cite{13}. Recently, there has been growing interest in the so-called DKP oscillator \cite{14,15,16,17,18,19,20,21,22,23} in particular in the background of a magnetic cosmic string \cite{13}. 
The cosmic strings and other topological defects can form at a cosmological phase transition\cite{24}. The conical nature of the space-time around the string causes a number of interesting physical effects. Until now, some problems have been investigated in the gravitational fields of topological defects including the one-electron atom problem.\cite{25,26,27}. Spinning cosmic strings similar usual cosmic string,  characterized by an angular parameter $  \alpha$ that depends on their linear mass density $  \mu$.
The DKP oscillator is described by performing the non-minimal coupling with a linear potential. The name distinguishes it from the system called a DKP oscillator with Lorentz tensor couplings of Refs. \cite{7,8,9,10,11,12,14,15,16}.
The DKP oscillator for spin-0 bosons has been investigated by Guo et al in Ref. \cite{10} in noncommutative phase space. The DKP oscillator with spin-0  hase been studied by Yang et al.  \cite{11}. Exact solution of DKP oscillator in the momentum space with the presence of minimal length has been analysed in\cite{8}. De Melo et al. construct the Galilean DKP equation for the harmonic oscillator in a non-commutative phase space.  \cite{28}. Falek and Merad investigated the DKP oscillator of spins 0 and 1 bosons in non- commutative space \cite{9}. Recently, there has been an increasing interest on the DKP oscillator \cite{13,14,15,16,28,29,30,31}. 
The non-relativistic limit of particle dynamics in curved space-time is considered in Refs. \cite{32,33,34,35,36}. Also, the dynamics of relativistic bosons and fermions in curved space-time is considered in Refs\cite{17,20,31}.\\
The influence of topological defect in the dynamics of bosons via DKP formalism has not been established for spinning cosmic strings. In this way, we consider the quantum dynamics of scalar bosons via DKP formalism embedded in the background of a spining cosmic string. 
We solve DKP equation in presence of the spinning cosmic string space time whose metric has off diagonal terms which involves time and space.
The influence of this topological defect in the energy spectrum and DKP spinor presented graphically.\\
The structure of this paper is as follows: Sec. 2 describes the covariant form of DKP equation in a spining cosmic string background. In Sec. 3, we introduce the DKP oscillator by performing the non-minimal coupling in this space time, and we obtain the radial equations that are solved. We plotted the DKP spinor, density of probability and the energy spectrum for different conditions involving the deficit angle, the oscillator frequency. In the Sect. 4 we present our conclusions.


\section{ Covariant form of the DKP equation in the spinning cosmic string background}

We choose the cosmic string spacetime background, where the line element is given by
\begin{align}\label{1}
ds^2=-dT^2+dX^2+dY^2+dZ^2
\end{align}
The spacetime generated by a spinning cosmic string without internal structure, which is termed ideal spinning cosmic string can be obtain by coordinate transformation as
\begin{align}\label{2}
\begin{gathered}
  T = t + a{\alpha ^{ - 1}}\varphi  \hfill \\
  X = rCos(\varphi ) \hfill \\
  Y = rSin(\varphi ) \hfill \\
  \varphi  = \alpha \varphi ' \hfill \\ 
\end{gathered} 
\end{align}

With this transformation, the line element (\eqref{1}) becomes\cite{37,38,39,40,41,42,43}

	\begin{align}\label{3}
	\begin{split}
	d{s^2} =  - {(dt + ad\varphi )^2} + d{r^2} + {\alpha ^2}{r^2}d{\varphi ^2} + d{z^2}\\	
	= - d{t^2} {\color{black}{+ d{r^2}}} - 2adtd\varphi  + ({\alpha ^2}{r^2} - {a^2})d{\varphi ^2} + d{z^2}{\text{ }}
	\end{split}
	\end{align}
	
with $- \infty  < z < \infty$ , $\rho \ge 0$ and $0 \le \varphi  \le 2\pi$. From this point on, we will take $c=1$. The angular parameter $ \alpha $runs in the interval $ \left( {0,1} \right] $  is related to the linear mass density $\mu$ of the string as $\alpha  = 1 - 4\mu$ and corresponds to a deficit angle $ \gamma=2\pi(1-\alpha) $. We take $a = 4Gj$ where $G $ is the universal gravitation constant and $j$ is the angular momentum of the spinning string; thus $a$ is a length that represents the rotation of the cosmic string. Note that in this case, the source of the gravitational field relative to a spinning cosmic string possesses angular momentum and the metric \eqref{1} has an off diagonal term involving time and space.

The DKP equation in the cosmic string spacetime\eqref{1} reads\cite{13,17,31}
\begin{align}\label{4}
({\rm{i}}{\beta ^\mu }\left( x \right){\nabla _\mu } - M)\Psi \left( x \right) = 0.
\end{align}
The covariant derivative in \eqref{4} is
\begin{align}\label{5}
{\nabla _\mu } = {\partial _\mu } + {\Gamma _\mu }\left( x \right)
\end{align}
Where ${\Gamma _\mu }$  are the spinorial affine connections given by
\begin{align}\label{6}
{\Gamma _\mu } = \frac{1}{2}{\omega _{\mu ab}}\left[ {{\beta ^a},{\beta ^b}} \right].
\end{align}
The matrices ${\beta ^a}$ are the standard Kemmer matrices in Minkowski spacetime. 
\begin{align}\label{7}
{\beta ^\mu } = e_a^\mu {\beta ^a}
\end{align}
The Kemmer matrices are an analogous to Dirac matrices in Dirac equation. There has been an increasing interest Dirac equation for spin half particles \cite{44,45,46,47}.
The matrices ${\beta ^a}$  satisfies the DKP algebra,
\begin{align}
{\beta ^\mu }{\beta ^\nu }{\beta ^\lambda } + {\beta ^\lambda }{\beta ^\nu }{\beta ^\mu } = {g^{\mu \nu }}{\beta ^\mu } + {g^{\lambda \nu }}{\beta ^\mu }.\nonumber
\end{align}
The conserved four-current is given by
\begin{align}
{J^\mu } = \frac{1}{2}\bar \Psi {\beta ^\mu }\Psi, \nonumber
\end{align}
and the conservation law for ${J^\mu }$ takes the form
\begin{align}\label{8}
{\nabla _\mu }{J^\mu } + \frac{i}{2}\bar \Psi (U - {\eta ^ 0 }{U^\dag }{\eta ^ 0 })\Psi  = \frac{1}{2}\bar \Psi ({\nabla _\mu }{\beta ^\mu })\Psi 
\end{align}
The adjoint spinor ${\bar \Psi }$ is defined as $\bar \Psi  = {\Psi ^\dag }{\eta ^ 0 }$ with ${\eta ^ 0 } = 2{\beta ^ 0 }{\beta ^ 0 } - 1$, in such a way that ${({\eta ^ 0 }{\beta ^\mu })^\dag } = {\eta ^ 0 }{\beta ^\mu }$. The factor $\frac{1}{2}$ which multiplies $\bar \Psi {\beta ^\mu }\Psi $, is of no importance for the conservation law, ensures the charge density is compatible with the one used in the Klein-Gordon theory and its non-relativistic limit. Thus, if $U$ is Hermitian with respect to $\eta ^ 0$ and the curved- space beta matrices are covariantly constant, then the four-current will be conserved if \cite{30}
\begin{align}
{\nabla _\mu }{\beta ^\mu } = 0.\nonumber
\end{align}
 The algebra expressed by these matrices generates a set of 126 independent matrices whose irreducible representations comprise a trivial representation, a five-dimensional representation describing the spin-zero particles and a ten-dimensional representation associated to spin-one particles. We choose the $5 \times 5$ beta-matrices as follows \cite{31}
\begin{align}\label{9}
 \begin{array}{l}
 {\beta ^0} = \left( {\begin{array}{*{20}{c}}
 	\theta &{{0_{2 \times 3}}}\\
 	{{0_{3 \times 2}}}&{{0_{3 \times 3}}}
 	\end{array}} \right),\vec \beta  = \left( {\begin{array}{*{20}{c}}
 	{{0_{2 \times 2}}}&{\vec \tau }\\
 	{ - {{\vec \tau }^T}}&{{0_{3 \times 3}}}
 	\end{array}} \right),\\
 \theta  = \left( {\begin{array}{*{20}{c}}
 	0&1\\
 	1&0
 	\end{array}} \right),{\tau  ^1} = \left( {\begin{array}{*{20}{c}}
 	{ - 1}&0&0\\
 	0&0&0
 	\end{array}} \right),{\tau ^2} = \left( {\begin{array}{*{20}{c}}
 	0&{ - 1}&0\\
 	0&0&0
 	\end{array}} \right),{\tau ^3} = \left( {\begin{array}{*{20}{c}}
 	0&0&{ - 1}\\
 	0&0&0
 	\end{array}} \right).
 \end{array}
\end{align}
In eq. \eqref{7}, $e_a^\mu $ denote the tetrad basis, that we can choose as 
\begin{align}\label{10}
e_a^\mu  = \left( {\begin{array}{*{20}{c}}
  1&{\frac{{a\operatorname{Sin} (\varphi )}}{{r\alpha }}}&{\frac{{ - a\operatorname{Cos} (\varphi )}}{{r\alpha }}}&0 \\ 
  0&{\operatorname{Cos} (\varphi )}&{\operatorname{Sin} (\varphi )}&0 \\ 
  0&{\frac{{ - \operatorname{Sin} (\varphi )}}{{r\alpha }}}&{\frac{{\operatorname{Cos} (\varphi )}}{{r\alpha }}}&0 \\ 
  0&0&0&1 
\end{array}} \right).
\end{align}
For the specific tetrad basis given by Eq. \eqref{10}, we find from Eq. \eqref{7} that the curved-space beta-matrices read
\begin{subequations}
	\begin{align}\label{11}
	\begin{gathered}
  {\beta ^{(0)}} = e_a^t{\beta ^a} = {\beta ^0} - \frac{a}{{r\alpha }}{\beta ^\varphi } \hfill \\
  {\beta ^{(1)}} = e_a^r{\beta ^a} = {\beta ^r} \hfill \\
  {\beta ^r} = \cos \varphi {\beta ^1} + \sin \varphi {\beta ^2} \hfill \\
  {\beta ^{(2)}} = e_a^2{\beta ^a} = \frac{{{\beta ^\varphi }}}{{r\alpha }} \hfill \\
  {\beta ^\varphi } =  - \sin \varphi {\beta ^1} + \cos \varphi {\beta ^2} \hfill \\
  {\beta ^{(3)}} = e_a^z{\beta ^a} = {\beta ^3} = {\beta ^z} \hfill \\ 
\end{gathered} 
	\end{align}
\end{subequations}
and the spin connections are given by
	\begin{align}\label{12}
	{\Gamma _\varphi} = (1 - \alpha)[\beta^1,\beta^2]
	\end{align}
We consider only the radial component in the non-minimal substitution.  Since the interaction is time-independent, one can write $\Psi \left( {r,t} \right) \propto {e^{im\varphi }}{e^{ik_zz}}{e^{-iEt}} \Phi (r)$, where $E$ is the energy of the scalar boson,  $m$ is the magnetic quantum number and $k_z$ is the wave number. The five-component DKP spinor can be written as ${\Phi ^T} = ({\Phi _1},{\Phi _2},{\Phi _3},{\Phi _4},{\Phi _5})$, and the DKP equation \eqref{4} leads to the five
\begin{align}\label{13}
\begin{gathered}
  (r\alpha \left( { - {\rm M}{\Phi _1}(r) + {\rm E}{\Phi _2}(r) + kz{\Phi _5}(r)} \right) + \cos \varphi \left( {\left( {a{\rm E} + m} \right){\Phi _4}(r) - {\text{i}}\left( {\left( { - 1 + \alpha } \right){\Phi _3}(r) + r\alpha {{\Phi '}_3}(r)} \right)} \right) -  \hfill \\
  \sin \varphi \left( {\left( {a{\rm E} + m} \right){\Phi _3}(r) + {\text{i}}\left( {\left( { - 1 + \alpha } \right){\Phi _4}(r) + r\alpha {{\Phi '}_4}(r)} \right)} \right)) = 0 \hfill \\
  \left( {{\rm E}{\Phi _1}(r) - {\rm M}{\Phi _2}(r)} \right) = 0 \hfill \\
  \left( {\left( {a{\rm E} + m} \right)\sin \varphi {\Phi _1}(r) + r\alpha \left( { - {\rm M}{\Phi _3}(r) + {\text{i}}\cos \phi {{\Phi '}_1}(r)} \right)} \right) = 0 \hfill \\
  \left( { - \left( {a{\rm E} + m} \right)\cos \varphi {\Phi _1}(r) + r\alpha \left( { - {\rm M}{\Phi _4}(r) + {\text{i}}\sin \varphi {{\Phi '}_1}(r)} \right)} \right) = 0 \hfill \\
  \left( {kz + {\rm M}\psi {\Phi _5}(r)} \right) = 0 \hfill \\ 
\end{gathered} 
\end{align}
Then we obtain the following equation of motion for the first component ${\Phi _1}$ of the DKP spinor :	
\begin{align}\label{14}
\left(-\frac{(a \text{E}+m )^2}{\alpha ^2 r^2}+\text{E}^2-\text{kz}^2-\text{M}^2\right){{{\Phi }_1}(r)}+\frac{(\alpha -1) {{{\Phi '}_1}(r)}}{\alpha  r}+{{\Phi ''}_1}(r)=0
\end{align}	
\begin{align}\label{15}
\Phi _1= {r^{ \frac{1}{2\alpha}}}{R_{n,\ell }}(r).
\end{align}
Then Eq. \eqref{14} changes to                                                                                                                                                                                                                                                                                                                                                                                                                                                                                                                                                                                                                                                                                                                                                                                                                                                                                                                                                                                                           
\begin{align}\label{16}
{{R''}_{n,m}}(r) + \frac{{{{R'}_{n,m}}(r)}}{r} + \left( {{E^2} - {k_z}^2 - {M^2} - \frac{{1 + 4{{\left( {aE + m} \right)}^2}}}{{4{r^2}{\alpha ^2}}}} \right)R(r) = 0
\end{align}

By the change of variable $ r=x\eta $, we can write Eq. \eqref{16} in the form

\begin{align}\label{17}
{{R''}_{n,m}}(x) + \frac{{{{R'}_{n,m}}(x)}}{x} + \left( {1 - \frac{{{\lambda ^2}}}{{{x^2}}}} \right)R(x) = 0
\end{align}
where $ \lambda = {(\frac{{1 + 4{{\left( {aE + m} \right)}^2}}}{{4{\alpha ^2}}})^{\frac{1}{2}}} $ and $ \eta  = {\left( {{E^2} - {k_z}^2 - {M^2}} \right)^{\frac{-1}{2}}} $.
The physical solution of  Eq.\eqref{17} are  $Bessel J $ function. Therefore the general solution to eq. (17) is given by\\
\begin{align}\label{18}
R(r) = {A_{\lambda \eta }}{J_\lambda }(\frac{r}{\eta }) + {B_{\lambda \eta }}{Y_\lambda }(\frac{r}{\eta })
\end{align}\\
where ${Y_\lambda }(\frac{r}{\eta })$ is the Bessel function of the second kind. Sometimes this family of functions is also called Neumann functions or Weber functions. ${J_\lambda }(\frac{r}{\eta })$ is the Bessel function of the first kind, given by\\
\begin{align}
{J_\nu }(x) = \sum\limits_{k = 0}^\infty  {\frac{{{{( - 1)}^k}}}{{k!\Gamma (k + \nu  + 1)}}} {\left( {\frac{x}{2}} \right)^{\nu  + 2k}}
\end{align}\\
and ${Y_\lambda }(\frac{r}{\eta })$ is the Bessel function of the second kind, given by\\
\begin{align}
{Y_\nu }(x) = \frac{{{J_\nu }(x)\operatorname{Cos} (\nu \pi ) - {J_{ - \nu }}(x)}}{{\operatorname{Sin} (\nu \pi )}}
\end{align}\\
By considering the boundary condition for Eq. \eqref{18} such that  ${B_{\lambda \eta }}= 0$ , we find\\
\begin{align}
R(r) = {A_{\lambda \eta }}{J_\lambda }(\frac{r}{\eta }) 
\end{align}\\
\section{ The DKP oscillator in spinning cosmic string background }

The DKP oscillator is introduced via the non-minimal substitution \cite{17,30,31}
\begin{align}
\frac{1}{i}{\vec \nabla _\alpha } \to \frac{1}{i}{\vec \nabla _\alpha } - iM\omega {\eta _ 0 }\vec r
\end{align}
where $\omega $ is the oscillator frequency, $M$ is the mass of the boson already found in Eq. \eqref{4}, and ${\vec \nabla }$ is defined in Eq. \eqref{5}.  We consider only the radial component in the non-minimal substitution. The DKP equation \eqref{4} leads to the five equations:
\begin{equation}\label{23}
\begin{gathered}
  (r\alpha \left( { - M{\Phi _1}(r) + E{\Phi _2}(r) + kz{\Phi _5}(r)} \right) + (\cos \varphi \left( {aE + m} \right){\Phi _4}(r)) +  \hfill \\
  ({\text{i}}\cos \varphi \left( {{\Phi _3}(r) + \alpha \left( { - 1 + {r^2}{\text{M}}\omega } \right){\Phi _3}(r) - r\alpha {{\Phi '}_3}(r)} \right)) -  \hfill \\
  \sin \varphi \left( {\left( {aE + m} \right){\Phi _3}(r) - {\text{i}}\left( {1 - \alpha  + {r^2}\alpha {\text{M}}\omega } \right){\Phi _4}(r) + {\text{i}}r\alpha {{\Phi '}_4}(r)} \right)) = 0 \hfill \\
  \left( {\left( {aE + m} \right)\sin \varphi {\Phi _1}(r) - Mr\alpha {\Phi _3}(r) + {\text{i}}r\alpha \cos \varphi \left( {r{\rm M}\omega {\Phi _1}(r) + {{\Phi '}_1}(r)} \right)} \right) = 0 \hfill \\
  \left( { - \left( {aE + m} \right)\cos \varphi {\Phi _1}(r) - Mr\alpha {\Phi _4}(r) + {\text{i}}r\alpha \sin \varphi \left( {r{\rm M}\omega {\Phi _1}(r) + {{\Phi '}_1}(r)} \right)} \right) = 0 \hfill \\
  \left( {E{\Phi _1}(r) - M{\Phi _2}(r)} \right) = 0 \hfill \\
  \left( {kz{\Phi _1}(r) + {\rm M}{\Phi _4}(r)} \right) = 0 \hfill \\ 
\end{gathered} 
\end{equation}
By solving the above system of equations \eqref{23} in favour of $\Phi_1$we get\\
\begin{align}
 \begin{gathered}
  {\Phi _2}(r) = \frac{E}{M}{\Phi _1}(r) \hfill \\
  {\Phi _5}(r) =  - \frac{{{k_z}}}{M}{\Phi _1}(r) \hfill \\
  {\Phi _4}(r) = \frac{{ - aECos\varphi {\Phi _1}(r) - m\operatorname{Sin} \varphi {\Phi _1}(r) + {\text{i}}\left( {{r^2}\alpha {\rm M}\omega \operatorname{Sin} \varphi {\Phi _1}(r) + r\alpha \operatorname{Sin} \varphi {{\Phi '}_1}(r)} \right)}}{{Mr\alpha }} \hfill \\
  {\Phi _3}(r) = \frac{{aE\operatorname{Sin} \varphi {\Phi _1}(r) + m\operatorname{Sin} \varphi {\Phi _1}(r) + {\text{i}}\left( {{r^2}\alpha {\rm M}\omega Cos\varphi {\Phi _1}(r) + r\alpha Cos\varphi {{\Phi '}_1}(r)} \right)}}{{Mr\alpha }} \hfill \\ 
\end{gathered} 
\end{align}

Combining these results we obtain an equation(20) of motion for the first component of the DKP spinor:
 \begin{align}\label{25}
 &{{\Phi ''}_1}(r) + \frac{{\left( { - 1 + \alpha } \right){{\Phi '}_1}(r)}}{{r\alpha }} \nonumber+\\
 & \left( {{E^2} - k{z^2} - {M^2} + 2M\omega  - \frac{{M\omega }}{\alpha } - \frac{{{{\left( {aE + m} \right)}^2}}}{{{r^2}{\alpha ^2}}} - {r^2}{{\rm M}^2}{\omega ^2}} \right){\Phi _1}(r) = 0
 \end{align}
Let us take $\Phi _1$as 
\begin{align}\label{26}
\Phi _1= {r^{ \frac{1}{2\alpha}}}{R_{n,\ell }}(r).
\end{align}
Then Eq. \eqref{25} changes to                                                                                                                                                                                                                                                                                                                                                                                                                                                                                                                                                                                                                                                                                                                                                                                                                                                                                                                                                                                                           
\begin{align}\label{}
&{R''_{n,m}}(r) + \frac{{{{R'}_{n,m}}(r)}}{r} \nonumber+\\
 &\left( {{E^2} - {k_Z}^2 - {M^2} + 2M\omega  - \frac{{M\omega }}{\alpha } - \frac{{1 + 4{{\left( {aE + m} \right)}^2}}}{{4{r^2}{\alpha ^2}}} - {r^2}{{\rm M}^2}{\omega ^2}} \right){R_{n,m}}(r) = 0
\end{align}
In order to solve the above equation, we employ the change of variable:
$s = r^2$, thus we rewrite the radial equation \eqref{27} in the form 
\begin{equation}\label{}
  {R_{n,m}}^{\prime \prime }(s) + \frac{1}{{s}}{R_{n,m}}^\prime (s) + \frac{1}{{{s^2}}}\left( { - \,\,{\xi _1}{s^2} + {\xi _2}s - {\xi _3}} \right){R_{n,m}}(r) = 0
\end{equation}
If we compare with this second-order differential equation with the Nikiforov-Uvarov (NU) form, given in Eq. \eqref{a.1} of Appendix A, we see that
\begin{align}\label{}
  \begin{gathered}
  {\xi _1} = \frac{{{M^2}{\omega ^2}}}{4} \hfill \\
  {\xi _2} = \frac{1}{4}({E^2} - k{z^2} - {M^2} + 2M\omega  - \frac{{M\omega }}{\alpha }), \hfill \\
  {\xi _3} = \frac{{1 - {\alpha ^2} + 4{{\left( {aE + m} \right)}^2}}}{{16{\alpha ^2}}} \hfill \\ 
\end{gathered} 
\end{align}
which gives the energy levels of the relativistic DKP equation from
\begin{align}\label{26}
(2n + 1)\sqrt {{\xi _1}}  - \xi _2+  2\sqrt {{\xi_3\xi_1}}  = 0,
\end{align}
where
\begin{align}\label{27}
\begin{array}{l}
{\alpha_1} = 1,{\alpha_2} = {\alpha_3} ={\alpha_4} ={\alpha_5} = 0 , {\alpha_6} = {\xi_1},\\
{\alpha_7} =  - {\xi_2},{\alpha_8} =  {\xi_3},{\alpha_9} = {\xi_1},{\alpha_{10}} =1+ 2\sqrt { {\xi_3}} ,{\alpha_{11}} = 2\sqrt {{ {\xi_1}}} \\
{\alpha_{12}} =  \sqrt {{ {\xi_3}}} ,{\alpha_{13}} =  - \sqrt {{ {\xi_1}}} 
\end{array}.
\end{align}
As the final step, it should be mentioned that the corresponding wave
function is
\begin{align}\label{28}
{R_{n,m}}(r) = {N r^{2{\alpha_{12}}}}{e^{{\alpha_{13}}{r^2}}}L_n^{{\alpha_{10}} - 1}({\alpha_{11}}{r^2}).
\end{align}
where $N$ is the normalization constant. {\color{black}{In limit 
	a$ \to$ 0 we have the usual metric in cylindrical coordinates where described by the line element
	\[d{s^2} =  - d{t^2} + d{r^2} + {\alpha ^2}{r^2}d{\varphi ^2} + d{z^2}\]
as pointed out by authors in \cite{17} dynamic of  DKP oscillator in the presence of  this metric describe by 
\[\begin{gathered}
  {{\varphi ''}_1}(r) + \frac{{\alpha  - 1}}{{\alpha r}}{{\varphi '}_1}(r) \hfill \\
   + ({E^2} - {M^2} - k{z^2} + \frac{{(2\alpha  - 1)M\omega }}{\alpha } - \frac{{{m^2}}}{{{\alpha ^2}{r^2}}} - {M^2}{\omega ^2}{r^2}){\varphi _1}(r) = 0 \hfill \\ 
\end{gathered} \]
and the  corresponding wave function is
\[\varphi (r) = N{r^{2A}}{e^{B{r^2}}}L_n^{C - 1}(D{r^2})\]
Were A,B,C and D are constant and  $L_n^{C-1 }$ denotes the generalized Laguerre polynomial.}}
In figure. 1, $ \Phi_1(r) $,  is plotted vs. $ r $ for different quantum number with the parameters listed  under it .The density of probability $ {\left| {{\Phi _1}} \right|^2}$ is shown in figure 2. The negative and positive solution of energy  vs. $ \alpha $ is shown in figure 3 and 4  for $n=1,5$ and 10. 
As in Fig. 3 and 4, we observe that the absolute value of energy  decreases with $ \alpha $. Also in figure. 5  energy  is plotted vs. $ \omega $ for  quantum numbers. We see that absolute value of energy  increases with $ \omega $. The negative and positive solution of energy  vs. $ n $ is shown in figure 6 for different parameter $ \alpha $. We obtained the energy levels of the DKP oscillator in that background and observed that the  energy increases with the level number.  In figures. 7, energy is plotted vs. $a$ for different quantum numbers. We see energy increases with parameter $a$.  Also We observed that the energy levels of the DKP oscillator in that background increases with the level number.

  \begin{figure}[H]
	\begin{center}
\includegraphics[scale=0.3]{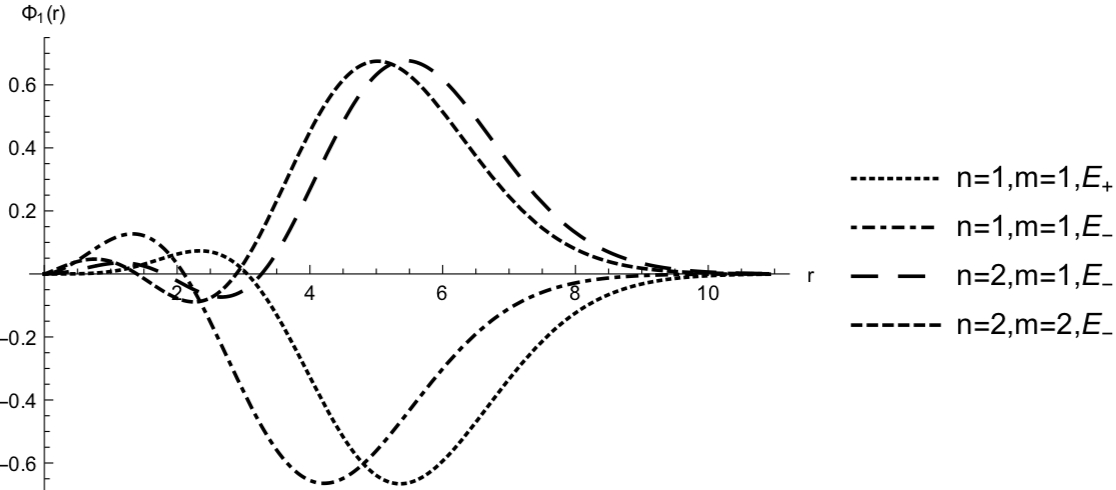}
\caption{The wave function $\Phi_1$ for $n=1,2$ and $0.0\leq r\leq 10.0$ GeV$^{-1}$, with the parameters $M=1$ GeV, $\alpha=0.9$, $\omega=0.25$ , $m=k_z=a=1$ }
  \end{center}
   \end{figure}

   \begin{figure}[H]
  \begin{center}
\includegraphics[scale=0.5]{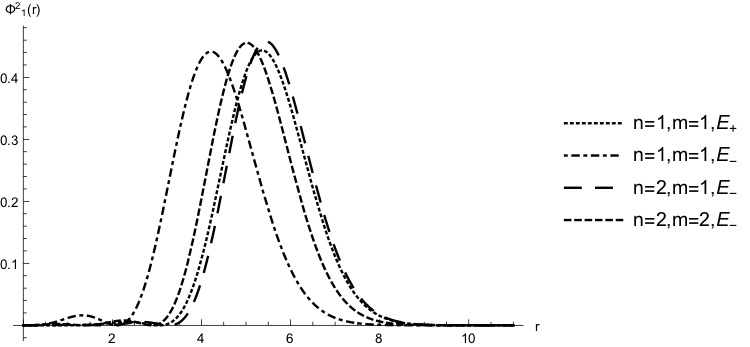}
\caption{Density of probability $\left|\Phi_1\right|^2$ for $n=1,2$ and $0.0\leq r\leq 10.0$ GeV$^{-1}$, with the parameters $M=1$ GeV, $\alpha=0.9$, $\omega=0.25$ , $m=k_z=a=1$}
  \end{center}
  
 \end{figure}
  
  \begin{figure}[H]
	\begin{center}
\includegraphics[scale=0.2]{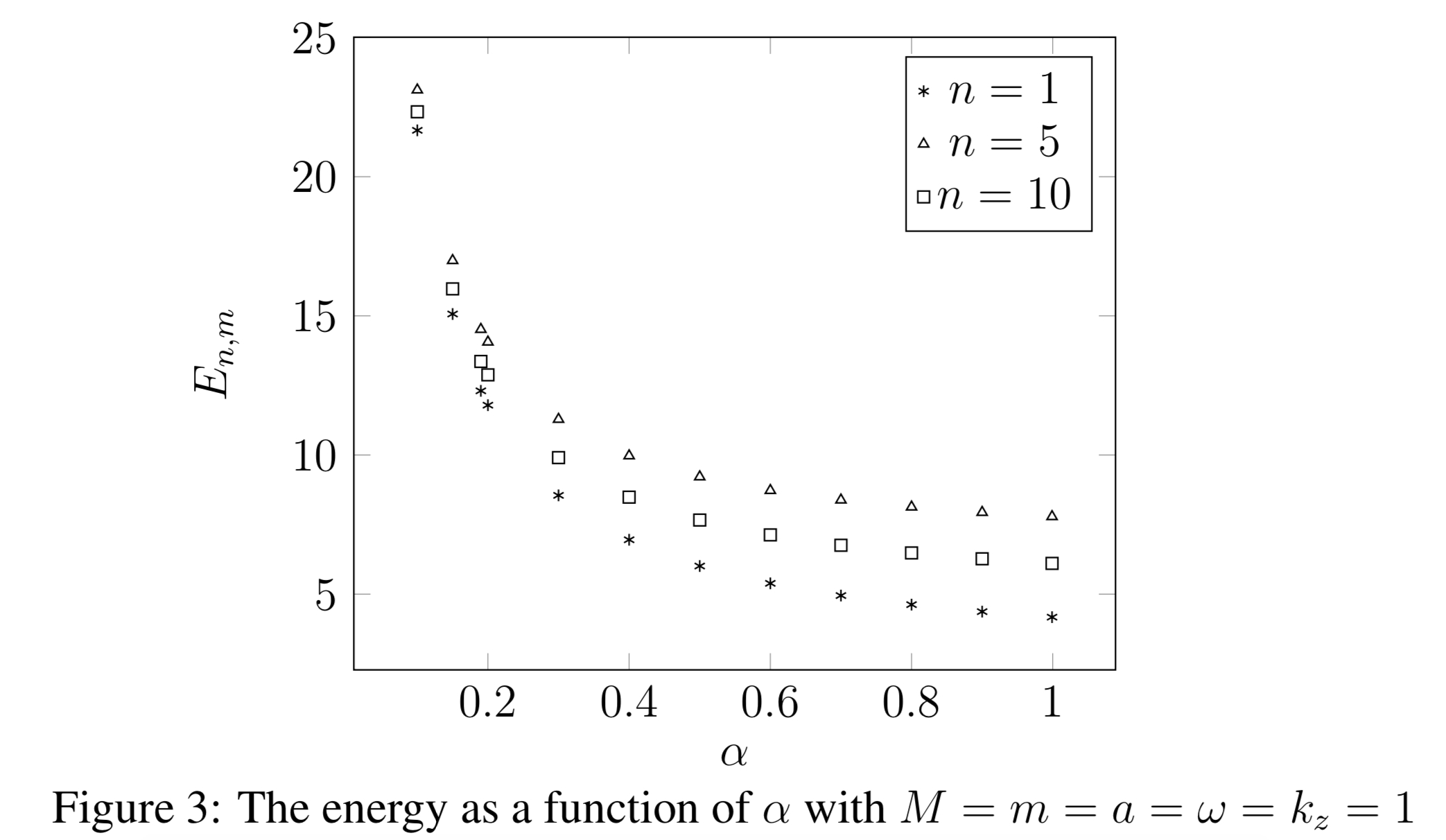}
  \end{center}
 \end{figure}
  
  \begin{figure}[H]
	\begin{center}
\includegraphics[scale=0.2]{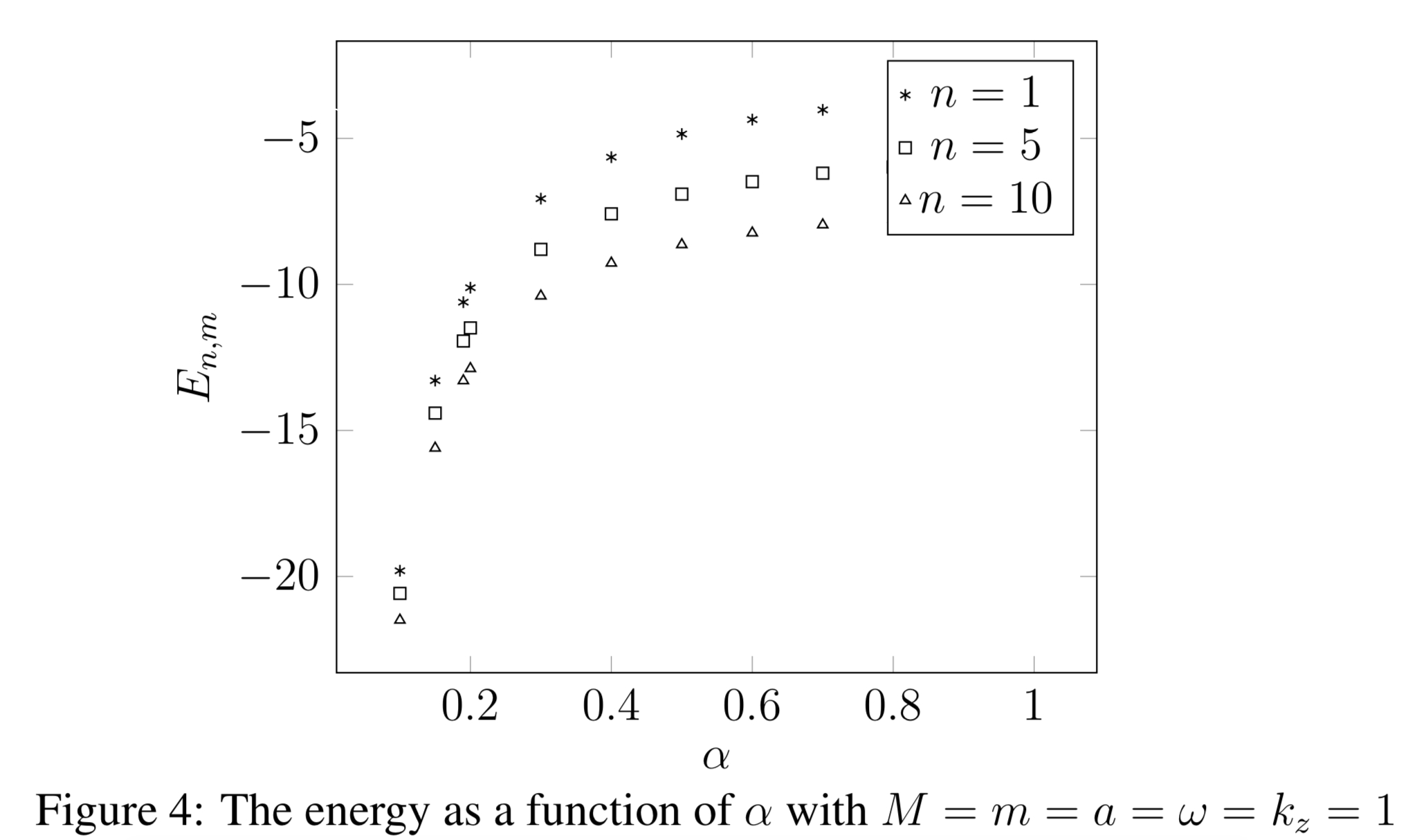}
  \end{center}
 \end{figure}
 
 \begin{figure}[H]
	\begin{center}
\includegraphics[scale=0.2]{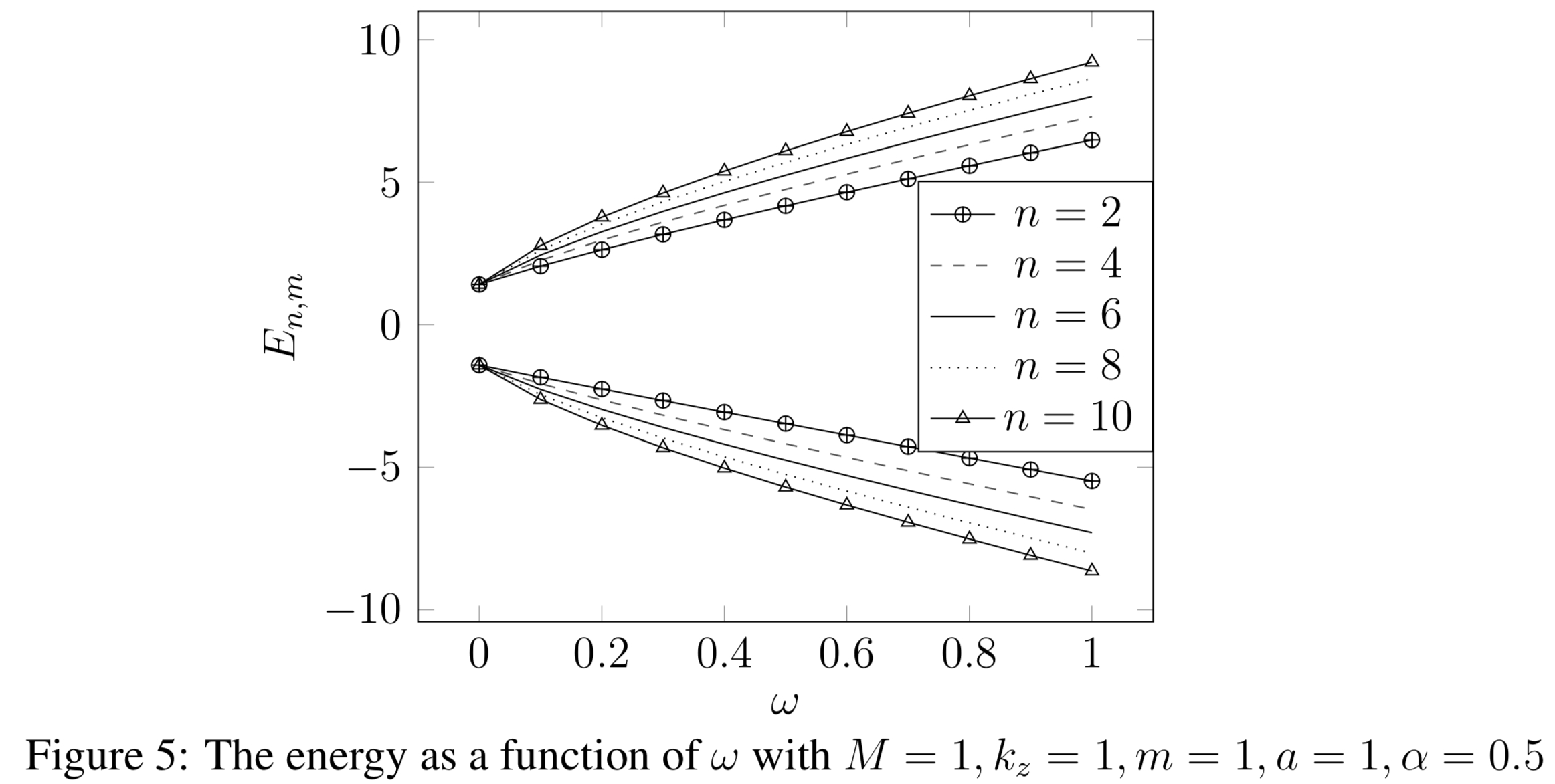}
  \end{center}
 \end{figure}
 
 \begin{figure}[H]
	\begin{center}
\includegraphics[scale=0.2]{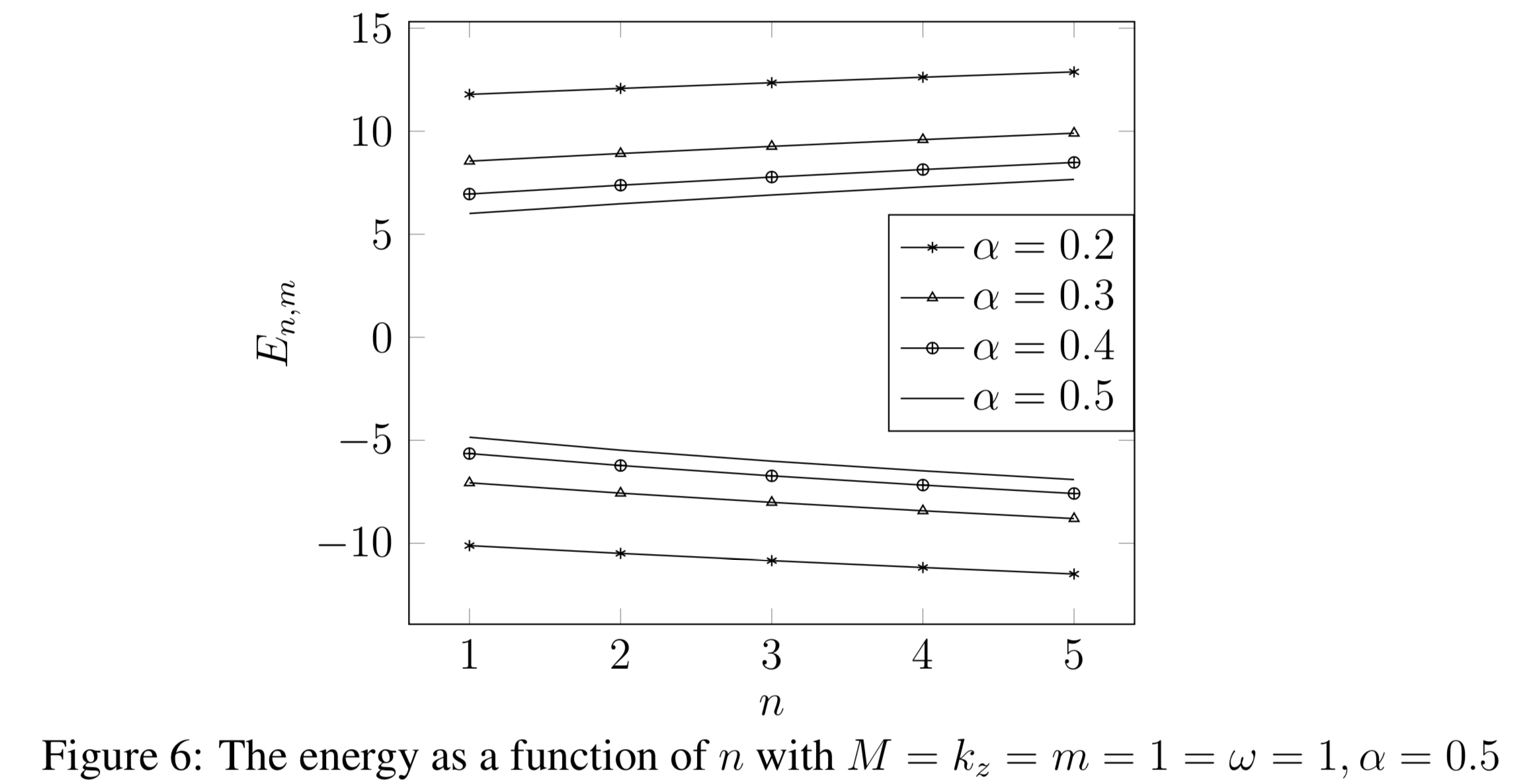}
  \end{center}
 \end{figure}

\begin{figure}[H]
	\begin{center}
\includegraphics[scale=0.2]{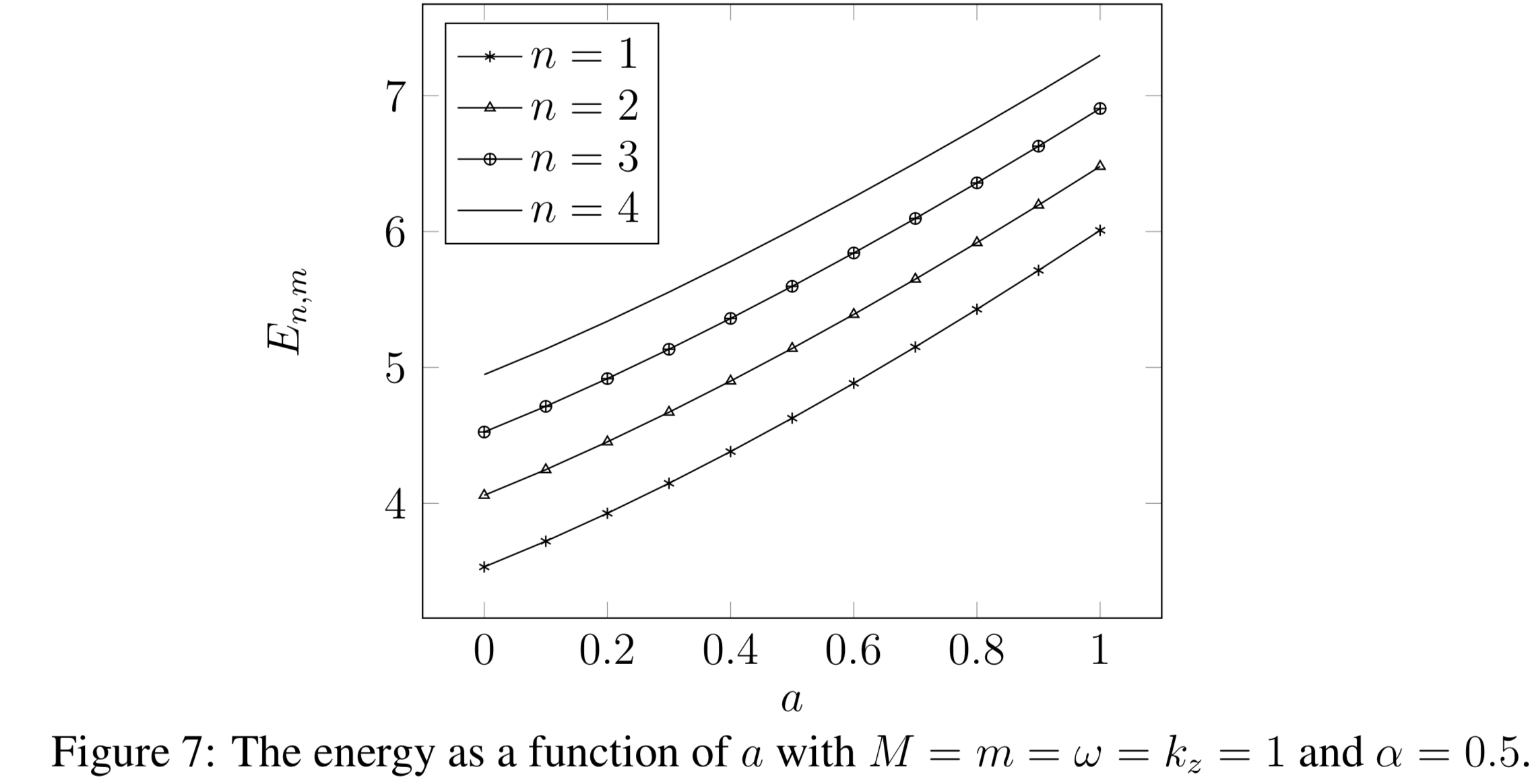}
  \end{center}
 \end{figure}

 \section{Conclusion}
 The overall objective of this paper is the study of the relativistic quantum dynamics of a DKP oscillator field  for spin-0 particle in the spinning cosmic string space-time. The line element in this background obtained by coordinate transformation of cartesian coordinate. The metric has off diagonal terms which involves time and space. We considered the covariant form of DKP equation in the spinning cosmic string background and obtained the solutions of DKP equation for spin-0 bosonse.
 Second we introduced DKP oscillator via the non-minimal substitution  and  considered DKP oscillator in that background.  From the corresponding DKP equation, we obtained a system of five equations. By combining the results of this system we obtained a second order differential equation for first component of DKP spinor that the solutions are Laguerre polynomials. We see the results are dependent on the linear mass density of the cosmic string. In the limit case of $a=0$ and $\alpha$ = 1, i.e., in the absence of a topological defect recover the general solution for flat space- time. We plotted $ \Phi_1(r) $, for $n=1,2$. We examined the behaviour of the density of probability $\left|\Phi_1\right|^2$ . We observed that $\left|\Phi_1\right|^2$ for any parameter by increasing $ r $ have a very small peak at beginning  and then have a taller peak and then by increasing $r$ it tend to zero.  
 We obtained the behaviour of energy spectrum as a function of $ \alpha $. We see that the absolute value of energy decreases  as $ \alpha $ increasing.\\\
{\color{black}{ \section*{Acknowledgment}
We thank the referee for a thorough reading of our manuscript and for constructive suggestions.}}
 \appendix
 \renewcommand{\theequation}{A.\arabic{equation}}
 \setcounter{equation}{0}
\section*{APPENDIX}
\section*{ Nikiforov-Uvarov (NU) method}
 The Nikiforov-Uvarov method is helpful in order to finde igenvalues and eigenfunctions of the Schr\"odinger equation, as well as other second-order differential equations of physical interest. More details can be found in Refs\cite{48,49}. According to this method, the eigenfunctions and eigenvalues of a second- order differential equation with potential
\begin{equation}\label{a.1}
  {\Phi ^{\prime \prime }}(s) + \frac{{{\alpha _1} - {\alpha _2}s}}{{s(1 - {\alpha _3}s)}}{\Phi ^\prime }(s) + \frac{1}{{{{(s(1 - {\alpha _3}s))}^2}}}\left( { - \,\,{\xi _1}{s^2} + {\xi _2}s - {\xi _3}} \right)\Phi (s) = 0
\end{equation}
According to the NU method, the eigenfunctions and eigenenergies, respectively, are 
  
\begin{align}\label{26}
\Phi (s) = {s^{{\alpha _{12}}}}{(1 - {\alpha _3}s)^{ - {\alpha _{12}} - ({{{\alpha _{13}}} \mathord{\left/
 {\vphantom {{{\alpha _{13}}} {{\alpha _3}}}} \right.
 \kern-\nulldelimiterspace} {{\alpha _3}}})}}P_n^{({\alpha _{10}} - 1,({{{\alpha _{11}}} \mathord{\left/
 {\vphantom {{{\alpha _{11}}} {{\alpha _3}}}} \right.
 \kern-\nulldelimiterspace} {{\alpha _3}}}) - {\alpha _{10}} - 1)}(1 - 2{\alpha _3}s)
 \end{align}
and
\begin{align}
{\alpha _2}n - (2n + 1){\alpha _5} + (2n + 1)(\sqrt {{\alpha _9}}  + {\alpha _3}\sqrt {{\alpha _8}} ) + n(n - 1){\alpha _3} + {\alpha _7} + 2{\alpha _3}{\alpha _8} + 2\sqrt {{\alpha _8}{\alpha _9}}  = 0
\end{align}\\
where
\begin{align}
 \begin{gathered}
  {\alpha _4} = \frac{1}{2}(1 - {\alpha _1}),\,\,\,{\alpha _5} = \frac{1}{2}({\alpha _2} - 2{\alpha _3}),\,\,\,\,{\alpha _6} = \alpha _5^2 + {\xi _1},\,\,\,\,\,\,{\alpha _7} = 2{\alpha _4}{\alpha _5} - {\xi _2}, \hfill \\
  {\alpha _8} = \alpha _4^2 + {\xi _3},\,\,\,\,\,{\alpha _9} = {\alpha _3}{\alpha _7} + \alpha _3^2{\alpha _8} + {\alpha _6},\,\,\,{\alpha _{10}} = {\alpha _1} + 2{\alpha _4} + 2\sqrt {{\alpha _8}} , \hfill \\
  {\alpha _{11}} = {\alpha _2} - 2{\alpha _5} + 2(\sqrt {{\alpha _9}}  + {\alpha _3}\sqrt {{\alpha _8}} ),\,\,\,{\alpha _{12}} = {\alpha _4} + \sqrt {{\alpha _8}} , \hfill \\
  {\alpha _{13}} = {\alpha _5} - (\sqrt {{\alpha _9}}  + {\alpha _3}\sqrt {{\alpha _8}} ) \hfill \\ 
\end{gathered} 
\end{align}

In the rather more special case of $\alpha =0$,
  \begin{align}
  \begin{gathered}
  \mathop {\lim }\limits_{{\alpha _3} \to 0} P_n^{({\alpha _{10}} - 1,({\alpha _{11}}/{\alpha _3}) - {\alpha _{10}} - 1)}(1 - 2{\alpha _3}s) = L_n^{{\alpha _{10}} - 1}({\alpha _{11}}s) \hfill \\
  \mathop {\lim }\limits_{{\alpha _3} \to 0} {(1 - {\alpha _3}s)^{ - {\alpha _{12}} - ({\alpha _{13}}/{\alpha _3})}} = {e^{{\alpha _{13}}s}} \hfill \\ 
\end{gathered} 
  \end{align}
and, from equation (11), we find for the wave function
\begin{align}
\Phi(s)  = {s^{{\alpha _{12}}}}{e^{{\alpha _{13}}s}}L_n^{{\alpha _{10}} - 1}({\alpha _{11}}s)
\end{align}
where $L_n^{\alpha_{10}-1 }$ denotes the generalized Laguerre polynomial.
 \\
 \\
 \\\\
 \\


\begin{thebibliography}{99}

\bibitem{1}  N. Kemmer, Proc. Roy. Soc., Ser. A 166, 127 (1938).
\bibitem{2}  R. J. Duffin, Phys. Rev. 54, 1114 (1938).
\bibitem{3}  N. Kemmer, Proc. Roy. Soc., Ser. A 173, 91 (1939).
\bibitem{4}  G. Petiau, Acad. Roy. Belg. Mem. Collect. 16, 1114 (1936).
\bibitem{5} E.M. Corson, Introduction to Tensors, Spinors Relativistic Wave Equations, Chelsea Pub. (1953).
\bibitem{6} W. Greiner, Relativistic Quantum Mechanics (Springer, Berlin, 2000).  
\bibitem{7}  M. Falek, M. Merad, J. Math. Phys. 50, 023508 (2009).
\bibitem{8}  M. Falek, M. Merad, J. Math. Phys. 51, 033516 (2010).
\bibitem{9}  M. Falek, M. Merad, Comm. Theor. Phys. 50, 587 (2008).
\bibitem{10}  G. Guo, C. Long, Z. Yang, S. Qin, Can. J. Phys. 87 989 (2009). 
\bibitem{11}  Z.H. Yang, C.Y. Long, S.J. Qin, Z.W. Long, Int. J. Theor. Phys. 49, 644 (2010).
\bibitem{12} H. Hassanabadi, Z. Molaee, S. Zarrinkamar, Eur. Phys. J. C 72, 2217 (2012).
\bibitem{13}  L.B. Castro, Eur. Phys. J. C 75, 287 (2015).
\bibitem{14}  N. Debergh, J. Ndimubandi, D. Strivay, Z. Phys. C 56, 421 (1992). 
\bibitem{15}  Y. Nedjadi, R.C. Barrett, J. Phys. A: Math. Gen. 27, 4301 (1994). 
\bibitem{16}  Y. Nedjadi, S. Ait-Tahar, R.C. Barrett, J. Phys. A: Math. Gen. 31, 3867 (1998).
\bibitem{17}  M.Hosseinpour,  H.Hassanabadi, F. M. Andrade. Eur. Phys. J. C (2018).
\bibitem{18}  A. Boumali, L. Chetouani, Phys. Lett. A 346, 261 (2005).
\bibitem{19}  I. Boztosun, M. Karakoc, F. Yasuk, A. Durmus, J. Math. Phys. 47, 062301 (2006). 
\bibitem{20}  M. de Montigny M. Hosseinpour and H. Hassanabadi. Int. J. Mod. Phys A. Vol. 31 (2016). 
\bibitem{21}  F. Yasuk, M. Karakoc, I. Boztosun, Phys. Scr. 78, 045010 (2008).
\bibitem{22} A. Boumali, J. Math. Phys. 49, 022302 (2008).
\bibitem{23}  Y. Kasri, L. Chetouani, Int. J. Theor. Phys. 47, 2249 (2008).
\bibitem{24}  A. Vilenkin, E.P.S. Shellard, Cosmic Strings and Other Topological Defects (Cambridge University Press, Cambridge, 1994); A. Vilenkin, Phys. Rep. 121, 263 (1985).
\bibitem{25} N. G. Marchuk, Nuov. Cim. B 115, 11 (2000).
\bibitem{26}  L. D. Landau, E. M. Lifshitz, Quantum Mechanics, Non-relativistic Theory (Pergamon, New York, 1977).
\bibitem{27}  G. de A. Marques, V. B. Bezerra, Phys. Rev. D 66, 105011 (2002).
\bibitem{28}. G.R. de Melo, M. de Montigny, E.S. Santos. J. Phys. Conf. Ser. 343, 012028 (2012).
\bibitem{29} A. Boumali. J. Math. Phys. 49(2), 022302 (2008).
\bibitem{30} L.B. Castro, Eur. Phys. J. C 76, 61 (2016).
\bibitem{31} H. Hassanabadi, M. Hosseinpour and M. de Montigny. Eur. Phys. J. Plus (2017)
\bibitem{32}  R. Bausch, R. Schmitz, L. A. Turski, Phys. Rev. Lett. 80, 2257 (1998).
\bibitem{33}  E. Aurell, J. Phys. A: Math. Gen. 32, 571 (1999).
\bibitem{34} C.R. Muniz, V.B. Bezerra, M.S. Cunha, Ann. Phys. 350, 105 (2014). 
\bibitem{35}  V.B. Bezerra, J. Math. Phys. 38, 2553 (1997).
\bibitem{36} C. Furtado, V. B. Bezerra, F. Moraes, Phys. Lett. A 289, 160 (2001).
\bibitem{37} M. S. Cunha et al,Eur. Phys. J. C (2016). 
\bibitem{38} G. Cl\'ement, Ann. Phys. (NY) 201, 241 (1990).
\bibitem{39} C.Furtado, F. Moraes and V. B. Bezerra ,Phys. Rev. D 59, 107504 (1999).
\bibitem{40} R. A. Puntigam and H. H. Soleng, Class. Quantum Grav. 14, 1129 (1997).
\bibitem{41} P.S.Letelier,Class. Quantum Grav. 12, 47 (1995).
\bibitem{42}  P.O. Mazur, Phys. Rev. Lett. 57, 8 (1986).
\bibitem{43} J. R. Gott and M. Alpert, Gen. Relativ. Gravitation 16, 243 (1984).
\bibitem{44} K Bakke ,Ann. Phys. (NY) 336, 489 (2013). 
\bibitem{45} G. Q. Garcia et al, Eur. Phys. J. Plus 132, 123 (2017).
\bibitem{46}  J Carvalho, Phys. Rev. A 84, 032109 (2011).
\bibitem{47} ERF Medeiros, Eur. Phys. J. C 72, 2051 (2012).
\bibitem{48} A.F. Nikiforov and V.B. Uvarov, Special Functions of Mathematical Physics, Birkh\"auser, Basel (1988).
\bibitem{49} C. Tezcan and R. Sever, Int. J. Theor. Phys. 48, 337 (2009).

\end{thebibliography}
\end{document}